\documentclass[prc,aps,twocolumn,showpacs,amssymb,superscriptaddress,float]{revtex4}

\usepackage{amsmath}
\usepackage{graphicx}
\usepackage{dcolumn}
\usepackage{float}

\begin{document}

\title{Shell-model calculations for neutron-rich carbon isotopes with
  a chiral nucleon-nucleon potential}

\author{L. Coraggio}
\affiliation{Istituto Nazionale di Fisica Nucleare, \\
Complesso Universitario di Monte  S. Angelo, Via Cintia - I-80126 Napoli,
Italy}
\author{A. Covello}
\affiliation{Istituto Nazionale di Fisica Nucleare, \\
Complesso Universitario di Monte  S. Angelo, Via Cintia - I-80126 Napoli,
Italy}
\affiliation{Dipartimento di Scienze Fisiche, Universit\`a
di Napoli Federico II, \\
Complesso Universitario di Monte  S. Angelo, Via Cintia - I-80126 Napoli,
Italy}
\author{A. Gargano}
\affiliation{Istituto Nazionale di Fisica Nucleare, \\
Complesso Universitario di Monte  S. Angelo, Via Cintia - I-80126 Napoli,
Italy}
\author{N. Itaco}
\affiliation{Istituto Nazionale di Fisica Nucleare, \\
Complesso Universitario di Monte  S. Angelo, Via Cintia - I-80126 Napoli,
Italy}
\affiliation{Dipartimento di Scienze Fisiche, Universit\`a
di Napoli Federico II, \\
Complesso Universitario di Monte  S. Angelo, Via Cintia - I-80126 Napoli,
Italy}

\date{\today}

\begin{abstract}
We have studied neutron-rich carbon isotopes in terms of the shell
model employing a realistic effective hamiltonian derived from the
chiral N$^3$LOW nucleon-nucleon potential. 
The single-particle energies and effective two-body interaction have
been both determined within the framework of the time-dependent
degenerate linked-diagram perturbation theory. 
The calculated results are in very good agreement with the available
experimental data, providing a sound description of this isotopic
chain toward the neutron dripline.
The correct location of the dripline is reproduced.
\end{abstract}


\maketitle
The advances in the production of exotic beams are making possible the
exploration of isotopes at and close to the neutron dripline.
This is a major goal of nuclear physics, since these studies may
reveal exotic features, such as shell modifications, that go beyond our
present understanding of the structure of nuclei.
In a recent Letter \cite{Tanaka10} Tanaka {\it et al.} have reported
on the observation of a large reaction cross section in the dripline
nucleus $^{22}$C.
The structure of $^{22}$C is quite interesting, because it can be
considered the heaviest Borromean nucleus ever observed
\cite{Kemper10}. 
This nucleus is weakly bound with a two-neutron separation energy
$S_{2n}$ evaluated to be $420 \pm 940$ keV \cite{Tanaka10}, and
may be seen as composed of three parts: two neutrons plus $^{20}$C.
These three pieces must all be present in order to obtain a bound 
$^{22}$C because of the particle instability of $^{21}$C.
This is the structure of a Borromean nucleus, which has only one bound
state but, considered as three-body system, admits no bound states
in the binary subsystems \cite{Zhukov93}.

The scope of the present work is to perform a theoretical study of the
properties of heavy carbon isotopes approaching the neutron dripline.
Our framework is the nuclear shell model, with both single-particle
energies and residual two-body interaction derived directly from a
realistic free nucleon-nucleon ($NN$) potential. 
This fully microscopic approach to shell-model calculations, where
no adjustable parameter is introduced, has already proved to be
successful in reproducing the neutron dripline for the oxygen isotopes
\cite{Coraggio07b} as well as the spectroscopic properties of the
$p$-shell nuclei \cite{Coraggio05c} and $N=82$ isotones
\cite{Coraggio09d}.

This is the first time that a realistic shell-model calculation is
performed in this region.
In fact, most of shell-model calculations to date have been performed
employing empirical single-particle energies and two-body matrix
elements adjusted to reproduce selected experimental data using the
full $psd$ model space (see for example \cite{Suzuki08b,Stanoiu08} and
references therein).

We now give a short description of our calculations.
We have carried out shell-model calculations, using the Oslo
shell-model code \cite{EngelandSMC}, for the heavy carbon isotopes in
terms of valence neutrons occupying the three single-particle levels
$0d_{5/2}$, $0d_{3/2}$, and $1s_{1/2}$, with $^{14}$C considered as an
inert core.
This choice is mainly motivated by the observed large energy gap
($\sim$ 6 MeV) between the ground and first excited state.
It is worth mentioning that in \cite{Ong09} it is suggested that the
anomalous suppression of the observed $B(E2;2^+_1 \rightarrow 0^+_1)$
transition rates in $^{14-18}$C indicates a possible proton-shell
closure in the neutron-rich carbon nuclei.

The shell-model effective hamiltonian $H_{\rm eff}$ has been derived
within the framework of the time-dependent degenerate linked-diagram
perturbation theory \cite{Coraggio09a}, starting from the N$^3$LOW
nucleon-nucleon potential \cite{Coraggio07b}.
The latter is a low-momentum potential derived from chiral perturbation
theory at next-to-next-to-next-to-leading order with a sharp momentum
cutoff at 2.1 fm$^{-1}$.
More explicitly, we have derived $H_{\rm eff}$ using the well-known
$\hat{Q}$-box plus folded-diagram method \cite{Coraggio09a}, where the
$\hat{Q}$-box is a collection of irreducible valence-linked Goldstone
diagrams, which we have calculated through third order in the N$^3$LOW
potential.

\begin{table}[H]
\caption{Theoretical shell-model single-particle energies (in MeV)
  employed in present work (see text for details).}
\begin{ruledtabular}
\begin{tabular}{cc}
$nlj$ & Single-particle energies \\
\colrule
$0d_{5/2}$  & 0.601 \\
$0d_{3/2}$  & 5.121 \\
$1s_{1/2}$  & -0.793 \\
\end{tabular}
\end{ruledtabular}
\label{spetab}
\end{table}

The hamiltonian $H_{\rm eff}$ contains one-body contributions, whose
collection is the so-called $\hat{S}$-box \cite{Shurpin83}. 
In realistic shell-model calculations it is customary to use a
subtraction procedure, so that only the two-body terms of $H_{\rm
  eff}$, which make up the effective interaction $V_{\rm eff}$, are
retained while the single-particle energies are taken from experiment.
In this work, however, we have adopted a different approach employing
single-particle energies obtained from the $\hat{S}$-box calculation.
In this connection, it is worth noting that the observed lowest
$J^{\pi}=\frac{1}{2}^+,~\frac{5}{2}^+,~\frac{3}{2}^+$ states in
$^{15}$C are not pure single-particle states \cite{Ajzenberg91}. 
In Table \ref{spetab} our calculated single-particle energies are
reported.
It should be noted that the experimental one-neutron separation energy
of $^{15}$C ($S_{n}$=1.22 MeV \cite{Audi03}) is slightly
underestimated.
For the sake of completeness, in Table \ref{tbme} we also report the
two-body matrix elements of $V_{\rm eff}$.

\begin{table}[H]
\caption{Proton-proton, neutron-neutron, and proton-neutron matrix
  elements (in MeV). They are antisymmetrized, and normalized by a
  factor $1/ \sqrt{ (1 + \delta_{j_aj_b})(1 + \delta_{j_cj_d})}$.}
\begin{ruledtabular}
\begin{tabular}{cccc}
$n_a l_a j_a ~ n_b l_b j_b ~ n_c l_c j_c ~ n_d l_d j_d $ & $J$ & $T_z$
  &  TBME \\
\colrule
 $ 0p_{ 1/2}~ 0p_{ 1/2}~ 0p_{ 1/2}~ 0p_{ 1/2}$ &  0 &  1 & -0.657 \\
 $ 0d_{ 5/2}~ 0d_{ 5/2}~ 0d_{ 5/2}~ 0d_{ 5/2}$ &  0 & -1 & -2.913 \\
 $ 0d_{ 5/2}~ 0d_{ 5/2}~ 0d_{ 3/2}~ 0d_{ 3/2}$ &  0 & -1 & -3.009 \\
 $ 0d_{ 5/2}~ 0d_{ 5/2}~ 1s_{ 1/2}~ 1s_{ 1/2}$ &  0 & -1 & -1.641 \\
 $ 0d_{ 3/2}~ 0d_{ 3/2}~ 0d_{ 3/2}~ 0d_{ 3/2}$ &  0 & -1 & -1.405 \\
 $ 0d_{ 3/2}~ 0d_{ 3/2}~ 1s_{ 1/2}~ 1s_{ 1/2}$ &  0 & -1 & -1.192 \\
 $ 1s_{ 1/2}~ 1s_{ 1/2}~ 1s_{ 1/2}~ 1s_{ 1/2}$ &  0 & -1 & -1.244 \\
 $ 0d_{ 5/2}~ 0d_{ 3/2}~ 0d_{ 5/2}~ 0d_{ 3/2}$ &  1 & -1 & -0.433 \\
 $ 0d_{ 5/2}~ 0d_{ 3/2}~ 0d_{ 3/2}~ 1s_{ 1/2}$ &  1 & -1 & -0.097 \\
 $ 0d_{ 3/2}~ 1s_{ 1/2}~ 0d_{ 3/2}~ 1s_{ 1/2}$ &  1 & -1 &  0.178 \\
 $ 0d_{ 5/2}~ 0d_{ 5/2}~ 0d_{ 5/2}~ 0d_{ 5/2}$ &  2 & -1 & -1.102 \\
 $ 0d_{ 5/2}~ 0d_{ 5/2}~ 0d_{ 5/2}~ 0d_{ 3/2}$ &  2 & -1 & -0.073 \\
 $ 0d_{ 5/2}~ 0d_{ 5/2}~ 0d_{ 5/2}~ 1s_{ 1/2}$ &  2 & -1 & -0.990 \\
 $ 0d_{ 5/2}~ 0d_{ 5/2}~ 0d_{ 3/2}~ 0d_{ 3/2}$ &  2 & -1 & -0.804 \\
 $ 0d_{ 5/2}~ 0d_{ 5/2}~ 0d_{ 3/2}~ 1s_{ 1/2}$ &  2 & -1 &  1.184 \\
 $ 0d_{ 5/2}~ 0d_{ 3/2}~ 0d_{ 5/2}~ 0d_{ 3/2}$ &  2 & -1 & -0.252 \\
 $ 0d_{ 5/2}~ 0d_{ 3/2}~ 0d_{ 5/2}~ 1s_{ 1/2}$ &  2 & -1 & -0.378 \\
 $ 0d_{ 5/2}~ 0d_{ 3/2}~ 0d_{ 3/2}~ 0d_{ 3/2}$ &  2 & -1 & -0.810 \\
 $ 0d_{ 5/2}~ 0d_{ 3/2}~ 0d_{ 3/2}~ 1s_{ 1/2}$ &  2 & -1 &  0.893 \\
 $ 0d_{ 5/2}~ 1s_{ 1/2}~ 0d_{ 5/2}~ 1s_{ 1/2}$ &  2 & -1 & -1.317 \\
 $ 0d_{ 5/2}~ 1s_{ 1/2}~ 0d_{ 3/2}~ 0d_{ 3/2}$ &  2 & -1 & -0.847 \\
 $ 0d_{ 5/2}~ 1s_{ 1/2}~ 0d_{ 3/2}~ 1s_{ 1/2}$ &  2 & -1 &  1.347 \\
 $ 0d_{ 3/2}~ 0d_{ 3/2}~ 0d_{ 3/2}~ 0d_{ 3/2}$ &  2 & -1 &  0.121 \\
 $ 0d_{ 3/2}~ 0d_{ 3/2}~ 0d_{ 3/2}~ 1s_{ 1/2}$ &  2 & -1 &  0.338 \\
 $ 0d_{ 3/2}~ 1s_{ 1/2}~ 0d_{ 3/2}~ 1s_{ 1/2}$ &  2 & -1 & -0.425 \\
 $ 0d_{ 5/2}~ 0d_{ 3/2}~ 0d_{ 5/2}~ 0d_{ 3/2}$ &  3 & -1 &  0.594 \\
 $ 0d_{ 5/2}~ 0d_{ 3/2}~ 0d_{ 5/2}~ 1s_{ 1/2}$ &  3 & -1 & -0.099 \\
 $ 0d_{ 5/2}~ 1s_{ 1/2}~ 0d_{ 5/2}~ 1s_{ 1/2}$ &  3 & -1 &  0.600 \\
 $ 0d_{ 5/2}~ 0d_{ 5/2}~ 0d_{ 5/2}~ 0d_{ 5/2}$ &  4 & -1 & -0.012 \\
 $ 0d_{ 5/2}~ 0d_{ 5/2}~ 0d_{ 5/2}~ 0d_{ 3/2}$ &  4 & -1 & -1.558 \\
 $ 0d_{ 5/2}~ 0d_{ 3/2}~ 0d_{ 5/2}~ 0d_{ 3/2}$ &  4 & -1 & -1.353 \\
 $ 0p_{ 1/2}~ 1s_{ 1/2}~ 0p_{ 1/2}~ 1s_{ 1/2}$ &  0 &  0 & -1.752 \\
 $ 0p_{ 1/2}~ 0d_{ 3/2}~ 0p_{ 1/2}~ 0d_{ 3/2}$ &  1 &  0 & -0.454 \\
 $ 0p_{ 1/2}~ 0d_{ 3/2}~ 0p_{ 1/2}~ 1s_{ 1/2}$ &  1 &  0 & -0.024 \\
 $ 0p_{ 1/2}~ 1s_{ 1/2}~ 0p_{ 1/2}~ 1s_{ 1/2}$ &  1 &  0 & -1.148 \\
 $ 0p_{ 1/2}~ 0d_{ 5/2}~ 0p_{ 1/2}~ 0d_{ 5/2}$ &  2 &  0 & -2.358 \\
 $ 0p_{ 1/2}~ 0d_{ 5/2}~ 0p_{ 1/2}~ 0d_{ 3/2}$ &  2 &  0 & -0.594 \\
 $ 0p_{ 1/2}~ 0d_{ 3/2}~ 0p_{ 1/2}~ 0d_{ 3/2}$ &  2 &  0 & -1.560 \\
 $ 0p_{ 1/2}~ 0d_{ 5/2}~ 0p_{ 1/2}~ 0d_{ 5/2}$ &  3 &  0 & -2.346 \\
\end{tabular}
\end{ruledtabular}
\end{table}

\begin{figure}[H]
\begin{center}
\includegraphics[scale=0.4,angle=0]{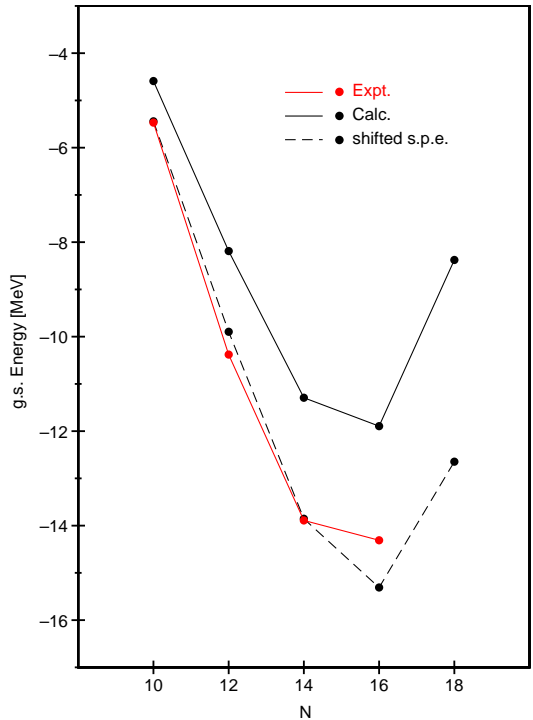}
\caption{(Color online)  Experimental \cite{Audi03,Tanaka10} and
  calculated ground-state  energies for carbon isotopes from $A=16$ to
  24. $N$ is the number of neutrons. See text for details.}
\label{evengse}
\end{center}
\end{figure}

We have performed calculations for carbon isotopes with $A$ ranging
from 16 up to 24, i.e. for systems with valence neutrons from $N_{\rm
  val}$=2 up to 10.
In Fig. \ref{evengse} the calculated ground-state energies of
even-mass isotopes (continuous black line) relative to $^{14}$C are
compared with the experimental ones (continuous red line)
\cite{Audi03}.
The experimental behavior is well reproduced, in particular our
results confirm that $^{22}$C is the last bound isotope; its
calculated $S_{2n}$ is 601 keV to be compared with the evaluation of
420 keV \cite{Audi03}.
Moreover, our calculations predict that $^{21}$C is unstable against
one-neutron decay, the theoretical $S_n$ being -1.6 MeV.
Therefore, our results fit the picture of $^{22}$C as a Borromean
nucleus.

From the inspection of Fig. \ref{evengse}, it can be seen that our
calculations underestimate the experimental data.
It is worth noting that this discrepancy may be ``healed'' by
downshifting the single-particle spectrum so as to reproduce the
experimental ground-state energy of $^{15}$C relative to $^{14}$C.
The results obtained with this downshift ($-427$ keV) are reported in
Fig. \ref{evengse} by the black dashed line.

\begin{figure}[H]
\begin{center}
\includegraphics[scale=0.35,angle=-90]{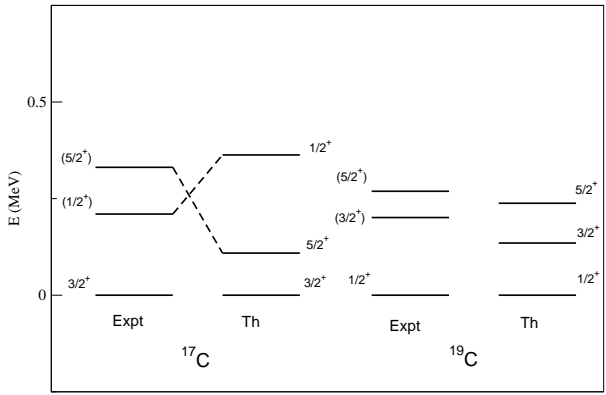}
\caption{Experimental \cite{Elekes05} and calculated low-energy
  spectra for $^{17}$C and $^{19}$C.}
\label{oddisotopes}
\end{center}
\end{figure}

In Fig. \ref{oddisotopes}, we report the experimental and calculated
low-energy levels of the odd-mass nuclei $^{17}$C and $^{19}$C.
It should be pointed out that the experimental levels shown in
Fig. \ref{oddisotopes} are the only observed bound states.
We see that the theory provides a satisfactory description of these
states, apart from the inversion between the $J^{\pi}=\frac{5}{2}^+$ and
$\frac{1}{2}^+$ states in $^{17}$C.

We have calculated the magnetic dipole moments of the
$J^{\pi}=(\frac{1}{2}^+)_1,~(\frac{5}{2}^+)_1$ states in $^{15}$C, and
$J^{\pi}=(\frac{3}{2}^+)_1$ state in $^{17}$C using an
effective operator obtained at third order in perturbation theory,
consistently with the derivation of $H_{\rm eff}$.
We have obtained $\mu(\frac{1}{2}^+_1)=-1.920$ n.m.,
$\mu(\frac{5}{2}^+_1)=-1.803$ n.m., and  $\mu(\frac{3}{2}^+_1)=-0.807$
n.m., to be compared with the experimental values $\pm 1.720 \pm
0.009$ \cite{Asahi02} n.m., $ \pm 1.758 \pm 0.030$
n.m. \cite{Asher80}, and $ \pm 0.758 \pm 0.004$ n.m. \cite{Ogawa02},
respectively.

Two magnetic dipole transition rates have been measured in $^{17}$C:
the $B(M1; \frac{1}{2}^+ \rightarrow \frac{3}{2}^+)$ and $B(M1;
\frac{5}{2}^+ \rightarrow \frac{3}{2}^+)$.
Our calculated values are respectively 0.0941 $\mu_{N}^2$ and 0.095
$\mu_{N}^2$, to be compared with the experimental values of $(0.0010
\pm 0.0001)~\mu_{N}^2$ and $(0.082~ ^{+0.032}_{-0.018})~\mu_{N}^2$
\cite{Suzuki08}.
It is worth mentioning that the anomalous quenching of the strength 
observed in the $\frac{1}{2}^+ \rightarrow \frac{3}{2}^+$ transition has 
been recently reproduced in Ref. \cite{Suzuki08b} in the framework of a 
shell-model calculation modifying the tensor components and the $T=1$
monopole terms of the empirical effective Hamiltonian $SFO$
\cite{Suzuki03} defined in the the full $psd$ model space.

\begin{table}[H]
\caption{Effective reduced single-neutron matrix elements of the
  magnetic dipole operator $M1$ (in n.m.).}
\begin{ruledtabular}
\begin{tabular}{cccc}
$n_a l_a j_a ~ n_b l_b j_b $ &  $\langle a || M1 || b \rangle $ \\
\colrule
 $0d_{ 5/2}~ 0d_{ 5/2}$  &   -2.553 \\
 $0d_{ 5/2}~ 0d_{ 3/2}$  &    2.743 \\
 $0d_{ 3/2}~ 0d_{ 5/2}$  &   -2.743 \\
 $0d_{ 3/2}~ 0d_{ 3/2}$  &    1.414 \\
 $0d_{ 3/2}~ 1s_{ 1/2}$  &    0.179 \\
 $1s_{ 1/2}~ 0d_{ 3/2}$  &   -0.179 \\
 $1s_{ 1/2}~ 1s_{ 1/2}$  &   -2.298 \\
\end{tabular}
\end{ruledtabular}
\label{tableM1effn}
\end{table}

In Tables \ref{tableM1effn} we report the effective reduced
single-neutron matrix elements of the $M1$ operator.

\begin{figure}[H]
\begin{center}
\includegraphics[scale=0.4,angle=90]{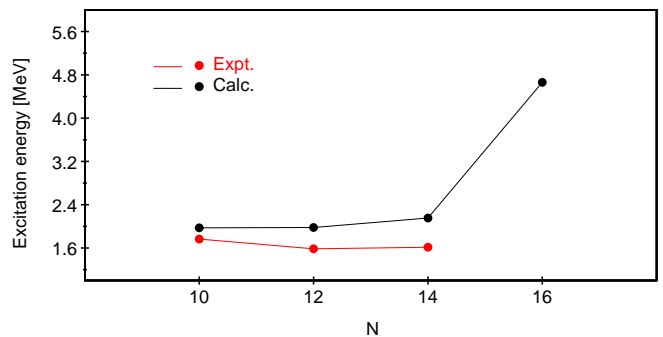}
\caption{(Color online) Experimental \cite{Ong08,Elekes09} and calculated
  excitation energies of the yrast $J^{\pi}=2^+$ states for carbon
  isotopes from $A=16$ to 22. $N$ is the number of neutrons.}
\label{J2p}
\end{center}
\end{figure}

In Fig. \ref{J2p} we report the experimental excitation energies of
the yrast $2^+$ states as a function of $A$ and compare them
with our calculated values.
It can be seen that the observed energies are reproduced nicely.
We also report our predicted excitation energy, 4.661 MeV, for the
unbound $J^{\pi}=2^+_1$ state in $^{22}$C.

\begin{table}[H]
\caption{Effective reduced single-neutron matrix elements of the
  electric quadrupole operator $E2$ (in ${\rm e~fm^2}$).}
\begin{ruledtabular}
\begin{tabular}{cccc}
$n_a l_a j_a ~ n_b l_b j_b $ &  $\langle a || E2 || b \rangle $ \\
\colrule
 $0d_{ 5/2}~ 0d_{ 5/2}$  &   -3.357 \\
 $0d_{ 5/2}~ 0d_{ 3/2}$  &   -1.866 \\
 $0d_{ 5/2}~ 1s_{ 1/2}$  &   -1.826 \\
 $0d_{ 3/2}~ 0d_{ 5/2}$  &    1.866 \\
 $0d_{ 3/2}~ 0d_{ 3/2}$  &   -2.803 \\
 $0d_{ 3/2}~ 1s_{ 1/2}$  &    1.434 \\
 $1s_{ 1/2}~ 0d_{ 5/2}$  &   -1.826 \\
 $1s_{ 1/2}~ 0d_{ 3/2}$  &   -1.434 \\
\end{tabular}
\end{ruledtabular}
\label{tableE2effn}
\end{table}

\begin{figure}[H]
\begin{center}
\includegraphics[scale=0.35,angle=-90]{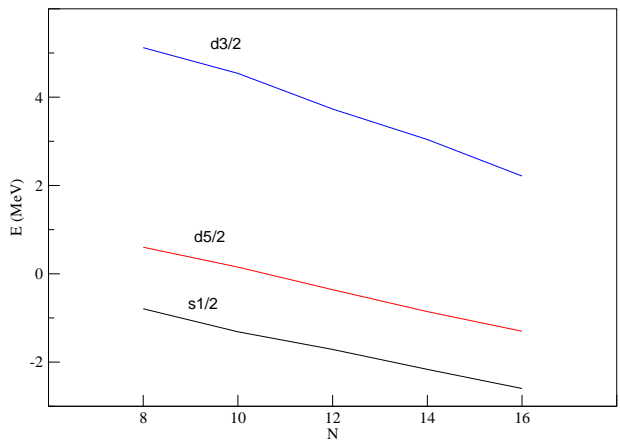}
\caption{(Color online) Effective single-particle energies of the
  neutron $1s_{1/2}$, $0d_{5/2}$, and $0d_{3/2}$ orbits from $A=16$ to
  22. $N$ is the number of neutrons.}
\label{figESPE}
\end{center}
\end{figure}

The behavior of the theoretical $2^+$ excitation energies confirms
that there is no $N=14$ subshell closure.
We obtain a ground-state wave function in $^{20}$C with only 14\%  of
the $(\nu d_{5/2})^6$ configuration (calculated using the OXBASH
shell-model code \cite{Oxbash}).
This is a direct consequence of the fact that the effective
single-particle energy \cite{Utsuno99} of the $1s_{1/2}$ state is the
lowest one all along the carbon isotopic chain, as can be seen in
Fig. \ref{figESPE}.
One can infer that the large energy gap between the $0d_{3/2}$ level
and the $0d_{5/2}$, $1s_{1/2}$ levels is responsible for the $N=16$
subshell closure.

It is worth mentioning that the behavior of the yrast $2^+$ state in
C isotopes is quite different from that shown by O isotopes, where a
$N=14$ subshell closure has been evidenced.
This is clearly related to the removal of the two protons from the
$0p_{1/2}$ level, which, in the case of oxygen isotopes, interact with
the $sd$ neutrons giving rise to a downshift of the $0d_{5/2}$
neutron level relative to the $1s_{1/2}$ one \cite{Stanoiu08}. 
In this connection, we have verified that, using our effective
hamiltonian,   the effective single particle energy of the $0d_{5/2}$
state in $^{17}$O is lowered by 2.1 MeV relative to that of the
$1s_{1/2}$ one.

We have calculated the $B(E2;2^+_1 \rightarrow 0^+_1)$ transition
rates up to $^{20}$C employing the effective operator of Table
\ref{tableE2effn}. 
Our results are compared with the experimental data in Table \ref{E2}. 
We see that the agreement is quite good, providing evidence for the
reliability of our calculated effective operator which takes into
account microscopically core-polarization effects up to third order in
the $NN$ potential.
It is well known \cite{Ong08,Ong09,Wiedeking08,Elekes09} that these
observed transition strengths are strongly hindered when compared with
the values obtained by an empirical formula based on the liquid-drop
model \cite{Raman88}.
In this connection, we would like to point out that our effective
reduced single-neutron matrix elements (see Table \ref{tableE2effn})
correspond to a neutron effective charge of about 0.4e with an
harmonic-oscillator parameter $b=1.72$ fm.

\begin{table}[H]
\caption{Experimental  and calculated reduced transition
  probabilities $B(E2;2^+_1 \rightarrow 0^+_1)$ (in e$^2$fm$^4$). 
  Superscripts $a,~b,~c$ refer to \cite{Ong09}, \cite{Wiedeking08},
  and \cite{Elekes09}, respectively. The experimental errors reported
  are the statistical and systematic ones, respectively.}
\begin{ruledtabular}
\begin{tabular}{ccc}
Nucleus &  Calc. & Expt. \\
\colrule
$^{16}$C & 1.8 & $2.6 \pm 0.2 \pm 0.7 ~^{\rm a}$ \\
     ~   & ~   & $4.15 \pm 0.73 ~^{\rm b}$ \\
$^{18}$C & 3.0 & $4.3 \pm 0.2 \pm 1.0 ~^{\rm a}$ \\
$^{20}$C & 3.7 & $ < 3.7~^{\rm c}$ \\
\end{tabular}
\end{ruledtabular}
\label{E2}
\end{table}

In summary, we have given here a shell-model description of heavy
carbon isotopes, using a fully microscopic approach.
This has been done by deriving from the realistic chiral $NN$
potential N$^3$LOW both the single-particle energies and residual
two-body interaction of the effective shell-model hamiltonian.
Our calculations have led to a sound description of these isotopes
when approaching the neutron dripline. 
As a matter of fact, they confirm the disappearance of the $N=14$
subshell closure that is present in the oxygen chain, and predict the
$N=16$ one.
In particular, we reproduce successfully the fact that $^{22}$C is the
last bound isotope.
It is worth mentioning that this nucleus, which has only recently been
observed \cite{Tanaka10}, is one of the most exotic of the 3000 known
isotopes, its $N/Z$ ratio being 2.67.
The quality of the results obtained makes us confident that our
approach may be a valuable tool to study nuclei far from the valley of
stability.

\bibliographystyle{apsrev}
\bibliography{biblio}

\begin{thebibliography}{26}
\expandafter\ifx\csname natexlab\endcsname\relax\def\natexlab#1{#1}\fi
\expandafter\ifx\csname bibnamefont\endcsname\relax
  \def\bibnamefont#1{#1}\fi
\expandafter\ifx\csname bibfnamefont\endcsname\relax
  \def\bibfnamefont#1{#1}\fi
\expandafter\ifx\csname citenamefont\endcsname\relax
  \def\citenamefont#1{#1}\fi
\expandafter\ifx\csname url\endcsname\relax
  \def\url#1{\texttt{#1}}\fi
\expandafter\ifx\csname urlprefix\endcsname\relax\def\urlprefix{URL }\fi
\providecommand{\bibinfo}[2]{#2}
\providecommand{\eprint}[2][]{\url{#2}}

\bibitem[{\citenamefont{Tanaka et~al.}(2010)\citenamefont{Tanaka, Yamaguchi,
  Suzuki, Ohtsubo, Fukuda, Nishimura, Takechi, Ogata, Ozawa, Izumikawa
  et~al.}}]{Tanaka10}
\bibinfo{author}{\bibfnamefont{K.}~\bibnamefont{Tanaka}},
  \bibinfo{author}{\bibfnamefont{T.}~\bibnamefont{Yamaguchi}},
  \bibinfo{author}{\bibfnamefont{T.}~\bibnamefont{Suzuki}},
  \bibinfo{author}{\bibfnamefont{T.}~\bibnamefont{Ohtsubo}},
  \bibinfo{author}{\bibfnamefont{M.}~\bibnamefont{Fukuda}},
  \bibinfo{author}{\bibfnamefont{D.}~\bibnamefont{Nishimura}},
  \bibinfo{author}{\bibfnamefont{M.}~\bibnamefont{Takechi}},
  \bibinfo{author}{\bibfnamefont{K.}~\bibnamefont{Ogata}},
  \bibinfo{author}{\bibfnamefont{A.}~\bibnamefont{Ozawa}},
  \bibinfo{author}{\bibfnamefont{T.}~\bibnamefont{Izumikawa}},
  \bibnamefont{et~al.}, \bibinfo{journal}{Phys. Rev. Lett.}
  \textbf{\bibinfo{volume}{104}}, \bibinfo{pages}{062701}
  (\bibinfo{year}{2010}).

\bibitem[{\citenamefont{Kemper and Cottle}(2010)}]{Kemper10}
\bibinfo{author}{\bibfnamefont{K.~W.} \bibnamefont{Kemper}} \bibnamefont{and}
  \bibinfo{author}{\bibfnamefont{P.~D.} \bibnamefont{Cottle}},
  \bibinfo{journal}{Physics} \textbf{\bibinfo{volume}{3}}, \bibinfo{eid}{13}
  (\bibinfo{year}{2010}).

\bibitem[{\citenamefont{Zhukov et~al.}(1993)\citenamefont{Zhukov, Danilin,
  Fedorov, Bang, Thompson, and Vaagen}}]{Zhukov93}
\bibinfo{author}{\bibfnamefont{M.~V.} \bibnamefont{Zhukov}},
  \bibinfo{author}{\bibfnamefont{B.~V.} \bibnamefont{Danilin}},
  \bibinfo{author}{\bibfnamefont{D.~V.} \bibnamefont{Fedorov}},
  \bibinfo{author}{\bibfnamefont{J.~M.} \bibnamefont{Bang}},
  \bibinfo{author}{\bibfnamefont{I.~J.} \bibnamefont{Thompson}},
  \bibnamefont{and} \bibinfo{author}{\bibfnamefont{J.~S.}
  \bibnamefont{Vaagen}}, \bibinfo{journal}{Phys. Rep.}
  \textbf{\bibinfo{volume}{231}}, \bibinfo{pages}{151} (\bibinfo{year}{1993}).

\bibitem[{\citenamefont{Coraggio et~al.}(2007)\citenamefont{Coraggio, Covello,
  Gargano, Itaco, Kuo, Entem, and Machleidt}}]{Coraggio07b}
\bibinfo{author}{\bibfnamefont{L.}~\bibnamefont{Coraggio}},
  \bibinfo{author}{\bibfnamefont{A.}~\bibnamefont{Covello}},
  \bibinfo{author}{\bibfnamefont{A.}~\bibnamefont{Gargano}},
  \bibinfo{author}{\bibfnamefont{N.}~\bibnamefont{Itaco}},
  \bibinfo{author}{\bibfnamefont{T.~T.~S.} \bibnamefont{Kuo}},
  \bibinfo{author}{\bibfnamefont{D.~R.} \bibnamefont{Entem}}, \bibnamefont{and}
  \bibinfo{author}{\bibfnamefont{R.}~\bibnamefont{Machleidt}},
  \bibinfo{journal}{Phys. Rev. C} \textbf{\bibinfo{volume}{75}},
  \bibinfo{pages}{024311} (\bibinfo{year}{2007}).

\bibitem[{\citenamefont{Coraggio and Itaco}(2005)}]{Coraggio05c}
\bibinfo{author}{\bibfnamefont{L.}~\bibnamefont{Coraggio}} \bibnamefont{and}
  \bibinfo{author}{\bibfnamefont{N.}~\bibnamefont{Itaco}},
  \bibinfo{journal}{Phys. Lett. B} \textbf{\bibinfo{volume}{616}},
  \bibinfo{pages}{43} (\bibinfo{year}{2005}).

\bibitem[{\citenamefont{Coraggio
  et~al.}(2009{\natexlab{a}})\citenamefont{Coraggio, Covello, Gargano, Itaco,
  and Kuo}}]{Coraggio09d}
\bibinfo{author}{\bibfnamefont{L.}~\bibnamefont{Coraggio}},
  \bibinfo{author}{\bibfnamefont{A.}~\bibnamefont{Covello}},
  \bibinfo{author}{\bibfnamefont{A.}~\bibnamefont{Gargano}},
  \bibinfo{author}{\bibfnamefont{N.}~\bibnamefont{Itaco}}, \bibnamefont{and}
  \bibinfo{author}{\bibfnamefont{T.~T.~S.} \bibnamefont{Kuo}},
  \bibinfo{journal}{Phys. Rev. C} \textbf{\bibinfo{volume}{80}},
  \bibinfo{pages}{044320} (\bibinfo{year}{2009}{\natexlab{a}}).

\bibitem[{\citenamefont{Suzuki and Otsuka}(2008)}]{Suzuki08b}
\bibinfo{author}{\bibfnamefont{T.}~\bibnamefont{Suzuki}} \bibnamefont{and}
  \bibinfo{author}{\bibfnamefont{T.}~\bibnamefont{Otsuka}},
  \bibinfo{journal}{Phys. Rev. C} \textbf{\bibinfo{volume}{78}},
  \bibinfo{pages}{061301(R)} (\bibinfo{year}{2008}).

\bibitem[{\citenamefont{Stanoiu et~al.}(2008)\citenamefont{Stanoiu, Sohler,
  Sorlin, Azaiez, Dombr\'adi, Brown, Belleguic, Borcea, Bourgeois, Dlouhy
  et~al.}}]{Stanoiu08}
\bibinfo{author}{\bibfnamefont{M.}~\bibnamefont{Stanoiu}},
  \bibinfo{author}{\bibfnamefont{D.}~\bibnamefont{Sohler}},
  \bibinfo{author}{\bibfnamefont{O.}~\bibnamefont{Sorlin}},
  \bibinfo{author}{\bibfnamefont{F.}~\bibnamefont{Azaiez}},
  \bibinfo{author}{\bibfnamefont{Z.}~\bibnamefont{Dombr\'adi}},
  \bibinfo{author}{\bibfnamefont{B.~A.} \bibnamefont{Brown}},
  \bibinfo{author}{\bibfnamefont{M.}~\bibnamefont{Belleguic}},
  \bibinfo{author}{\bibfnamefont{C.}~\bibnamefont{Borcea}},
  \bibinfo{author}{\bibfnamefont{C.}~\bibnamefont{Bourgeois}},
  \bibinfo{author}{\bibfnamefont{Z.}~\bibnamefont{Dlouhy}},
  \bibnamefont{et~al.}, \bibinfo{journal}{Phys. Rev. C}
  \textbf{\bibinfo{volume}{78}}, \bibinfo{pages}{034315}
  (\bibinfo{year}{2008}).

\bibitem[{\citenamefont{Engeland}()}]{EngelandSMC}
\bibinfo{author}{\bibfnamefont{T.}~\bibnamefont{Engeland}}, \bibinfo{note}{the
  Oslo shell-model code 1991-2006, unpublished}.

\bibitem[{\citenamefont{Ong et~al.}(2009)\citenamefont{Ong, Imai, Suzuki,
  Iwasaki, Sakurai, Onishi, Suzuki, Ota, Takeuchi, Nakao et~al.}}]{Ong09}
\bibinfo{author}{\bibfnamefont{H.~J.} \bibnamefont{Ong}},
  \bibinfo{author}{\bibfnamefont{N.}~\bibnamefont{Imai}},
  \bibinfo{author}{\bibfnamefont{D.}~\bibnamefont{Suzuki}},
  \bibinfo{author}{\bibfnamefont{H.}~\bibnamefont{Iwasaki}},
  \bibinfo{author}{\bibfnamefont{H.}~\bibnamefont{Sakurai}},
  \bibinfo{author}{\bibfnamefont{T.~K.} \bibnamefont{Onishi}},
  \bibinfo{author}{\bibfnamefont{M.~K.} \bibnamefont{Suzuki}},
  \bibinfo{author}{\bibfnamefont{S.}~\bibnamefont{Ota}},
  \bibinfo{author}{\bibfnamefont{S.}~\bibnamefont{Takeuchi}},
  \bibinfo{author}{\bibfnamefont{T.}~\bibnamefont{Nakao}},
  \bibnamefont{et~al.}, \bibinfo{journal}{Eur. Phys. J. A}
  \textbf{\bibinfo{volume}{42}}, \bibinfo{pages}{393} (\bibinfo{year}{2009}).

\bibitem[{\citenamefont{Coraggio
  et~al.}(2009{\natexlab{b}})\citenamefont{Coraggio, Covello, Gargano, Itaco,
  and Kuo}}]{Coraggio09a}
\bibinfo{author}{\bibfnamefont{L.}~\bibnamefont{Coraggio}},
  \bibinfo{author}{\bibfnamefont{A.}~\bibnamefont{Covello}},
  \bibinfo{author}{\bibfnamefont{A.}~\bibnamefont{Gargano}},
  \bibinfo{author}{\bibfnamefont{N.}~\bibnamefont{Itaco}}, \bibnamefont{and}
  \bibinfo{author}{\bibfnamefont{T.~T.~S.} \bibnamefont{Kuo}},
  \bibinfo{journal}{Prog. Part. Nucl. Phys.} \textbf{\bibinfo{volume}{62}},
  \bibinfo{pages}{135} (\bibinfo{year}{2009}{\natexlab{b}}).

\bibitem[{\citenamefont{Shurpin et~al.}(1983)\citenamefont{Shurpin, Kuo, and
  Strottman}}]{Shurpin83}
\bibinfo{author}{\bibfnamefont{J.}~\bibnamefont{Shurpin}},
  \bibinfo{author}{\bibfnamefont{T.~T.~S.} \bibnamefont{Kuo}},
  \bibnamefont{and}
  \bibinfo{author}{\bibfnamefont{D.}~\bibnamefont{Strottman}},
  \bibinfo{journal}{Nucl. Phys. A} \textbf{\bibinfo{volume}{408}},
  \bibinfo{pages}{310} (\bibinfo{year}{1983}).

\bibitem[{\citenamefont{Ajzenberg-Selove}(1991)}]{Ajzenberg91}
\bibinfo{author}{\bibfnamefont{F.}~\bibnamefont{Ajzenberg-Selove}},
  \bibinfo{journal}{Nucl. Phys. A} \textbf{\bibinfo{volume}{523}},
  \bibinfo{pages}{1} (\bibinfo{year}{1991}).

\bibitem[{\citenamefont{Audi et~al.}(2003)\citenamefont{Audi, Wapstra, and
  Thibault}}]{Audi03}
\bibinfo{author}{\bibfnamefont{G.}~\bibnamefont{Audi}},
  \bibinfo{author}{\bibfnamefont{A.~H.} \bibnamefont{Wapstra}},
  \bibnamefont{and} \bibinfo{author}{\bibfnamefont{C.}~\bibnamefont{Thibault}},
  \bibinfo{journal}{Nucl. Phys. A} \textbf{\bibinfo{volume}{729}},
  \bibinfo{pages}{337} (\bibinfo{year}{2003}).

\bibitem[{\citenamefont{Asahi et~al.}(2002)\citenamefont{Asahi, Sakai, Ogawa,
  Miyoshi, Yogo, Goto, Suga, Ueno, Kobayashi, Yoshimi et~al.}}]{Asahi02}
\bibinfo{author}{\bibfnamefont{K.}~\bibnamefont{Asahi}},
  \bibinfo{author}{\bibfnamefont{K.}~\bibnamefont{Sakai}},
  \bibinfo{author}{\bibfnamefont{H.}~\bibnamefont{Ogawa}},
  \bibinfo{author}{\bibfnamefont{H.}~\bibnamefont{Miyoshi}},
  \bibinfo{author}{\bibfnamefont{K.}~\bibnamefont{Yogo}},
  \bibinfo{author}{\bibfnamefont{A.}~\bibnamefont{Goto}},
  \bibinfo{author}{\bibfnamefont{T.}~\bibnamefont{Suga}},
  \bibinfo{author}{\bibfnamefont{H.}~\bibnamefont{Ueno}},
  \bibinfo{author}{\bibfnamefont{Y.}~\bibnamefont{Kobayashi}},
  \bibinfo{author}{\bibfnamefont{A.}~\bibnamefont{Yoshimi}},
  \bibnamefont{et~al.}, \bibinfo{journal}{Nucl. Phys. A}
  \textbf{\bibinfo{volume}{704}}, \bibinfo{pages}{88c} (\bibinfo{year}{2002}).

\bibitem[{\citenamefont{Asher et~al.}(1980)\citenamefont{Asher, Bennett, Brown,
  Doubt, and Grace}}]{Asher80}
\bibinfo{author}{\bibfnamefont{J.}~\bibnamefont{Asher}},
  \bibinfo{author}{\bibfnamefont{D.~W.} \bibnamefont{Bennett}},
  \bibinfo{author}{\bibfnamefont{B.~A.} \bibnamefont{Brown}},
  \bibinfo{author}{\bibfnamefont{H.~A.} \bibnamefont{Doubt}}, \bibnamefont{and}
  \bibinfo{author}{\bibfnamefont{M.~A.} \bibnamefont{Grace}},
  \bibinfo{journal}{J. Phys. G} \textbf{\bibinfo{volume}{6}},
  \bibinfo{pages}{251} (\bibinfo{year}{1980}).

\bibitem[{\citenamefont{Ogawa et~al.}(2002)\citenamefont{Ogawa, Asahi, Ueno,
  Sakai, Miyoshi, Kameda, Suzuki, Izumi, Imai, Watanabe et~al.}}]{Ogawa02}
\bibinfo{author}{\bibfnamefont{H.}~\bibnamefont{Ogawa}},
  \bibinfo{author}{\bibfnamefont{K.}~\bibnamefont{Asahi}},
  \bibinfo{author}{\bibfnamefont{H.}~\bibnamefont{Ueno}},
  \bibinfo{author}{\bibfnamefont{K.}~\bibnamefont{Sakai}},
  \bibinfo{author}{\bibfnamefont{H.}~\bibnamefont{Miyoshi}},
  \bibinfo{author}{\bibfnamefont{D.}~\bibnamefont{Kameda}},
  \bibinfo{author}{\bibfnamefont{T.}~\bibnamefont{Suzuki}},
  \bibinfo{author}{\bibfnamefont{H.}~\bibnamefont{Izumi}},
  \bibinfo{author}{\bibfnamefont{N.}~\bibnamefont{Imai}},
  \bibinfo{author}{\bibfnamefont{Y.~X.} \bibnamefont{Watanabe}},
  \bibnamefont{et~al.}, \bibinfo{journal}{Eur. Phys. J. A}
  \textbf{\bibinfo{volume}{13}}, \bibinfo{pages}{81} (\bibinfo{year}{2002}).

\bibitem[{\citenamefont{Elekes et~al.}(2005)\citenamefont{Elekes, Dombr{\'a}di,
  Kanungo, Baba, F{\"u}l{\"o}p, Gibelin, Horv{\'a}th, Ideguchi, Ichikawa, Iwasa
  et~al.}}]{Elekes05}
\bibinfo{author}{\bibfnamefont{Z.}~\bibnamefont{Elekes}},
  \bibinfo{author}{\bibfnamefont{Z.}~\bibnamefont{Dombr{\'a}di}},
  \bibinfo{author}{\bibfnamefont{R.}~\bibnamefont{Kanungo}},
  \bibinfo{author}{\bibfnamefont{H.}~\bibnamefont{Baba}},
  \bibinfo{author}{\bibfnamefont{Z.}~\bibnamefont{F{\"u}l{\"o}p}},
  \bibinfo{author}{\bibfnamefont{J.}~\bibnamefont{Gibelin}},
  \bibinfo{author}{\bibfnamefont{{\'A}.}~\bibnamefont{Horv{\'a}th}},
  \bibinfo{author}{\bibfnamefont{E.}~\bibnamefont{Ideguchi}},
  \bibinfo{author}{\bibfnamefont{Y.}~\bibnamefont{Ichikawa}},
  \bibinfo{author}{\bibfnamefont{N.}~\bibnamefont{Iwasa}},
  \bibnamefont{et~al.}, \bibinfo{journal}{Phys. Lett. B}
  \textbf{\bibinfo{volume}{614}}, \bibinfo{pages}{174} (\bibinfo{year}{2005}).

\bibitem[{\citenamefont{Suzuki et~al.}(2008)\citenamefont{Suzuki, Iwasaki, Ong,
  Imai, Sakurai, Nakao, Aoi, Baba, Bishop, Ichikawa et~al.}}]{Suzuki08}
\bibinfo{author}{\bibfnamefont{D.}~\bibnamefont{Suzuki}},
  \bibinfo{author}{\bibfnamefont{H.}~\bibnamefont{Iwasaki}},
  \bibinfo{author}{\bibfnamefont{H.~J.} \bibnamefont{Ong}},
  \bibinfo{author}{\bibfnamefont{N.}~\bibnamefont{Imai}},
  \bibinfo{author}{\bibfnamefont{H.}~\bibnamefont{Sakurai}},
  \bibinfo{author}{\bibfnamefont{T.}~\bibnamefont{Nakao}},
  \bibinfo{author}{\bibfnamefont{N.}~\bibnamefont{Aoi}},
  \bibinfo{author}{\bibfnamefont{H.}~\bibnamefont{Baba}},
  \bibinfo{author}{\bibfnamefont{S.}~\bibnamefont{Bishop}},
  \bibinfo{author}{\bibfnamefont{Y.}~\bibnamefont{Ichikawa}},
  \bibnamefont{et~al.}, \bibinfo{journal}{Phys. Lett. B}
  \textbf{\bibinfo{volume}{666}}, \bibinfo{pages}{222} (\bibinfo{year}{2008}).

\bibitem[{\citenamefont{Suzuki et~al.}(2003)\citenamefont{Suzuki, Fujimoto, and
  Otsuka}}]{Suzuki03}
\bibinfo{author}{\bibfnamefont{T.}~\bibnamefont{Suzuki}},
  \bibinfo{author}{\bibfnamefont{R.}~\bibnamefont{Fujimoto}}, \bibnamefont{and}
  \bibinfo{author}{\bibfnamefont{T.}~\bibnamefont{Otsuka}},
  \bibinfo{journal}{Phys. Rev. C} \textbf{\bibinfo{volume}{67}},
  \bibinfo{pages}{044302} (\bibinfo{year}{2003}).

\bibitem[{\citenamefont{Ong et~al.}(2008)\citenamefont{Ong, Imai, Suzuki,
  Iwasaki, Sakurai, Onishi, Suzuki, Ota, Takeuchi, Nakao et~al.}}]{Ong08}
\bibinfo{author}{\bibfnamefont{H.~J.} \bibnamefont{Ong}},
  \bibinfo{author}{\bibfnamefont{N.}~\bibnamefont{Imai}},
  \bibinfo{author}{\bibfnamefont{D.}~\bibnamefont{Suzuki}},
  \bibinfo{author}{\bibfnamefont{H.}~\bibnamefont{Iwasaki}},
  \bibinfo{author}{\bibfnamefont{H.}~\bibnamefont{Sakurai}},
  \bibinfo{author}{\bibfnamefont{T.~K.} \bibnamefont{Onishi}},
  \bibinfo{author}{\bibfnamefont{M.~K.} \bibnamefont{Suzuki}},
  \bibinfo{author}{\bibfnamefont{S.}~\bibnamefont{Ota}},
  \bibinfo{author}{\bibfnamefont{S.}~\bibnamefont{Takeuchi}},
  \bibinfo{author}{\bibfnamefont{T.}~\bibnamefont{Nakao}},
  \bibnamefont{et~al.}, \bibinfo{journal}{Phys. Rev. C}
  \textbf{\bibinfo{volume}{78}}, \bibinfo{pages}{014308}
  (\bibinfo{year}{2008}).

\bibitem[{\citenamefont{Elekes et~al.}(2009)\citenamefont{Elekes, Dombr\'adi,
  Aiba, Aoi, Baba, Bemmerer, Brown, Furumoto, F\"ul\"op, Iwasa
  et~al.}}]{Elekes09}
\bibinfo{author}{\bibfnamefont{Z.}~\bibnamefont{Elekes}},
  \bibinfo{author}{\bibfnamefont{Z.}~\bibnamefont{Dombr\'adi}},
  \bibinfo{author}{\bibfnamefont{T.}~\bibnamefont{Aiba}},
  \bibinfo{author}{\bibfnamefont{N.}~\bibnamefont{Aoi}},
  \bibinfo{author}{\bibfnamefont{H.}~\bibnamefont{Baba}},
  \bibinfo{author}{\bibfnamefont{D.}~\bibnamefont{Bemmerer}},
  \bibinfo{author}{\bibfnamefont{B.~A.} \bibnamefont{Brown}},
  \bibinfo{author}{\bibfnamefont{T.}~\bibnamefont{Furumoto}},
  \bibinfo{author}{\bibfnamefont{Z.}~\bibnamefont{F\"ul\"op}},
  \bibinfo{author}{\bibfnamefont{N.}~\bibnamefont{Iwasa}},
  \bibnamefont{et~al.}, \bibinfo{journal}{Phys. Rev. C}
  \textbf{\bibinfo{volume}{79}}, \bibinfo{pages}{011302}
  (\bibinfo{year}{2009}).

\bibitem[{\citenamefont{Brown et~al.}()\citenamefont{Brown, Etchegoyen, and
  Rae}}]{Oxbash}
\bibinfo{author}{\bibfnamefont{B.~A.} \bibnamefont{Brown}},
  \bibinfo{author}{\bibfnamefont{A.}~\bibnamefont{Etchegoyen}},
  \bibnamefont{and} \bibinfo{author}{\bibfnamefont{W.~D.~M.}
  \bibnamefont{Rae}}, \bibinfo{note}{the Computer Code OXBASH, MSU-NSCL Report
  No. 524}.

\bibitem[{\citenamefont{Utsuno et~al.}(1999)\citenamefont{Utsuno, Otsuka,
  Mizusaki, and Honma}}]{Utsuno99}
\bibinfo{author}{\bibfnamefont{Y.}~\bibnamefont{Utsuno}},
  \bibinfo{author}{\bibfnamefont{T.}~\bibnamefont{Otsuka}},
  \bibinfo{author}{\bibfnamefont{T.}~\bibnamefont{Mizusaki}}, \bibnamefont{and}
  \bibinfo{author}{\bibfnamefont{M.}~\bibnamefont{Honma}},
  \bibinfo{journal}{Phys. Rev. C} \textbf{\bibinfo{volume}{60}},
  \bibinfo{pages}{054315} (\bibinfo{year}{1999}).

\bibitem[{\citenamefont{Wiedeking et~al.}(2008)\citenamefont{Wiedeking, Fallon,
  Macchiavelli, Gibelin, Basunia, Clark, Cromaz, Deleplanque, Gros, Jeppesen
  et~al.}}]{Wiedeking08}
\bibinfo{author}{\bibfnamefont{M.}~\bibnamefont{Wiedeking}},
  \bibinfo{author}{\bibfnamefont{P.}~\bibnamefont{Fallon}},
  \bibinfo{author}{\bibfnamefont{A.~O.} \bibnamefont{Macchiavelli}},
  \bibinfo{author}{\bibfnamefont{J.}~\bibnamefont{Gibelin}},
  \bibinfo{author}{\bibfnamefont{M.~S.} \bibnamefont{Basunia}},
  \bibinfo{author}{\bibfnamefont{R.~M.} \bibnamefont{Clark}},
  \bibinfo{author}{\bibfnamefont{M.}~\bibnamefont{Cromaz}},
  \bibinfo{author}{\bibfnamefont{M.-A.} \bibnamefont{Deleplanque}},
  \bibinfo{author}{\bibfnamefont{S.}~\bibnamefont{Gros}},
  \bibinfo{author}{\bibfnamefont{H.~B.} \bibnamefont{Jeppesen}},
  \bibnamefont{et~al.}, \bibinfo{journal}{Phys. Rev. Lett.}
  \textbf{\bibinfo{volume}{100}}, \bibinfo{pages}{152501}
  (\bibinfo{year}{2008}).

\bibitem[{\citenamefont{Raman et~al.}(1988)\citenamefont{Raman, Nestor, and
  Bhatt}}]{Raman88}
\bibinfo{author}{\bibfnamefont{S.}~\bibnamefont{Raman}},
  \bibinfo{author}{\bibfnamefont{C.~W.} \bibnamefont{Nestor}},
  \bibnamefont{and} \bibinfo{author}{\bibfnamefont{K.~H.} \bibnamefont{Bhatt}},
  \bibinfo{journal}{Phys. Rev. C} \textbf{\bibinfo{volume}{37}},
  \bibinfo{pages}{805} (\bibinfo{year}{1988}).

\end{thebibliography}

\end{document}